\documentclass[notitlepage,aps,floats,showpacs,showkeys,preprintnumbers,nofootinbib]{revtex4-1}

\usepackage{amsmath,amssymb}
\usepackage{epsfig}

\def\lsim{\mathrel{\raise.3ex\hbox{$<$\kern-.75em\lower1ex\hbox{$\sim$}}}}
\def\gsim{\mathrel{\raise.3ex\hbox{$>$\kern-.75em\lower1ex\hbox{$\sim$}}}}

\newcommand{\nc}{\newcommand}
\nc{\beq}{\begin{equation}}  \nc{\eeq}{\end{equation}}
\nc{\bea}{\begin{eqnarray}}  \nc{\eea}{\end{eqnarray}}
\nc{\baa}{\begin{array}}     \nc{\eaa}{\end{array}}
\nc{\bit}{\begin{itemize}}   \nc{\eit}{\end{itemize}}
\nc{\ben}{\begin{enumerate}} \nc{\een}{\end{enumerate}}
\nc{\bce}{\begin{center}}    \nc{\ece}{\end{center}}
\nc{\bpm}{\begin{pmatrix}}   \nc{\epm}{\end{pmatrix}}
\nc{\bvt}{\begin{verbatim}}  \nc{\evt}{\end{verbatim}}

\nc{\bew}{\beta_W}
\nc{\dtz}{d^2_{0,0}}
\nc{\dzz}{d^0_{0,0}}
\nc{\dto}{d^2_{1,0}}
\def\kap{\kappa}

\def\tev{~{\rm TeV}}
\def\gev{~{\rm GeV}}

\def\bea{\begin{eqnarray}}
\def\eea{\end{eqnarray}}

\def\beq{\begin{equation}}
\def\eeq{\end{equation}}
\def\eg{{\it e.g.}}
\def\ie{{\it i.e.}}

\def\bce{\begin{center}}
\def\ece{\end{center}}

\def\ww{W^+W^-}

\def\br{{\rm BR}}

\def\hsm{h_{SM}}
\def\mhsm{m_{\hsm}}

\def\Hhat{\widehat H}

\def\phio{\phi_0}

\def\ie{{\it i.e.}}
\def\anti{\overline}

\def\mw{m_W}

\def\gam{\gamma}

\def\mh{m_{h}}
\def\mphi{m_\phi}
\def\what{\widehat}
\def\lwh{\widehat\Lambda_W}

\def\lphi{\Lambda_\phi}

\def\mphi{m_\phi}

\def\hbar{\overline h}

\def\mpl{M_{Pl}}

\def\call{{\cal L}}

\def\mpl{m_{Pl}}
\def\mompl{m_0/\mpl}

\def\ww{W^+W^-}

\def\half{{\small 1\over 2}}
\def\mbt{{\half m_0 b_0}}
\def\gh{g_h}
\def\gphi{g_\phi}
\def\ghr{\gh^r}
\def\gphir{\gphi^r}
\def\Eq#1{Eq.~(\ref{#1})}
\def\bfbm{\bf \boldmath}

\def\monekk{m_1^{\rm KK}}

\begin{document}
\title{Higgs-Radion interpretation of the LHC data?}
\author{Bohdan Grzadkowski}
\affiliation{Institute of Theoretical Physics,  \\
   Faculty of Physics, University of Warsaw, 00-681 Warsaw, Poland}

\author{John F. Gunion}
\affiliation{ Department of  Physics \\
University of California, Davis, CA 95616, U.S.A.}

\author{Manuel Toharia}
\affiliation{Department of Physics, Concordia University\\
7141 Sherbrooke St. West, Montreal\\ Quebec, CANADA H4B 1R6}

\begin{abstract}
  We explore the parameter choices in the five-dimensional
  Randall-Sundrum model with the inclusion of Higgs-radion mixing that
  can describe current LHC hints for one or more Higgs boson signals.

\end{abstract}

\maketitle

\vspace*{-.3in}
\section{Introduction}
\label{intro}
\vspace*{-.1in}

The two simplest ways of reconciling the weak energy scale ${\cal
  O}(1\tev)$ and the much higher GUT or reduced Planck mass scale
${\cal O}(10^{18}\gev)$ in a consistent theory are (i) to employ
supersymmetry or (ii) to introduce one or more warped extra
dimensions.  In this letter, we pursue the  5D version of the latter
introduced by Randall and Sundrum~(RS) ~\cite{Randall:1999ee}, but
modified in that all fields other than the Higgs reside in the
bulk. Having the gauge and fermion fields in the bulk is needed to
adequately suppress flavor changing neutral current (FCNC)  operators
and operators contributing to precision electroweak (PEW) corrections
~\cite{Davoudiasl:1999tf,Pomarol:1999ad,Gherghetta:2000qt,Davoudiasl:2000wi,Csaki:2002gy,Hewett:2002fe,Agashe:2003zs,Agashe:2007zd}.   

In the notation of \cite{Dominici:2002jv}, the background RS metric that solves Einstein's equations takes the form
\beq
ds^2=e^{-2m_0b_0 |y|} \eta_{\mu\nu} dx^\mu dx^\nu -b_0^2dy^2
\eeq
where $y$ is the coordinate for the 5th dimension with $|y|\leq 1/2$.
%
The graviton and radion fields, $h_{\mu\nu}(x,y)$ and $\phi_0(x)$, are
the quantum fluctuations relative to the background metric
$\eta_{\mu\nu}$ and $b_0$, respectively.  In particular, $\phio(x)$ is
the quantum degree of freedom associated with fluctuations of the
distance between the branes. In the simplest case, only gravity
propagates in the bulk while the SM is located on the infrared (or
TeV) brane at $y=1/2$ and the interactions of Kaluza-Klein (KK)
gravitons and the radion with the SM are described by 
\beq
\call_{\rm int}=-{1\over\lwh}\sum_{n\neq 0} h_{\mu\nu}^n T^{\mu\nu} -{\phi_0\over\lphi}T_\mu^\mu
\label{int}
\eeq
where $h_{\mu\nu}^n(x)$ are the KK modes (with mass
$m_n$) of the graviton field $h_{\mu\nu}(x,y)$.  In the above,  $\lwh \simeq \sqrt 2
\mpl \Omega_0$, where $\Omega_0 = e^{-\mbt}$, and $\lphi = \sqrt
3\,\lwh$ is the vacuum expectation value of the radion field.  Note from Eq.~(\ref{int}) that the
radion couples to matter with coupling strength $1/\lphi$. 
If matter and gauge fields propagate in the bulk then the interactions
of gravitons and the radion with the matter and gauge fields are
controlled by the overlap of appropriate 5th-dimensional profiles and
corrections to \Eq{int} appear.  

In addition to the radion, the model contains a conventional Higgs boson,
$h_0$.  The RS model provides a simple solution to the hierarchy
problem if the Higgs is placed on the TeV brane at $y=1/2$ by virtue
of the fact that the 4D electro-weak scale $v_0$ is given in terms of the 
${\cal  O}(\mpl)$  5D Higgs vev, $\what v$, by: 
%
$v_0= \Omega_0 \what v=
e^{-\mbt} \what v \sim 1\tev$ for $\mbt \sim
35\,.  
$
%
As a result, to solve the hierarchy problem, $\lphi=\sqrt 6 \mpl\Omega_0$ should not exceed a few TeV~\cite{Randall:1999ee}. 

The ratio $\mompl$ is a particularly crucial parameter that
characterizes the 5-dimensional curvature.
As discussed shortly, large curvature values $\mompl\gsim 0.5$ are
favored for fitting the LHC Higgs excesses and by bounds on FCNC and
PEW constraints. In early discussions of the RS model it was argued
that $R_5/M_5^2<1$ ($M_5$ being the  5D Planck scale and $R_5=20m_0^2 $
the size of the  5D curvature) is needed to suppress higher curvature
terms in the  5D action, which leads to $\mompl\lsim 0.15$ being
preferred.  However \cite{Agashe:2007zd} argues that $R_5/\Lambda^2$
(with $\Lambda$ being the energy scale at which the  5D gravity theory
becomes strongly coupled, estimated by naive dimensional analysis to
be $\Lambda\sim 2\sqrt 3 \pi M_5$) is the appropriate measure,
implying that values as large as $\mompl<\sqrt{3\pi^3/(5\sqrt 5)}\sim
3$ are acceptable. In this regard, the relation between the mass of
the 1st KK graviton excitation ($G^1$), $\mompl$ and $\lphi$, 
\beq
\monekk={(\mompl)  x_1^{\rm KK}\over\sqrt 6} \lphi\,,
\label{monekk}
\eeq
where $x_1^{\rm KK}\sim 3.83$ is the 1st zero of the Bessel function
$J_1$, will require large $\mompl$ if the lower bound on $m_1^{\rm KK}
$ is large and $\lphi\sim 1\tev$. 

In the simplest RS scenario, the SM fermions and gauge bosons are
confined to the brane.  However, this is now regarded as highly
problematical because the higher-dimensional operators in the 5D
effective field theory are suppressed only by $\tev^{-1}$ and then
FCNC processes and PEW observable corrections are predicted to be much
too large.  This arrangement also provides no explanation of the
flavor hierarchies.  It is therefore now regarded as necessary
~\cite{Davoudiasl:1999tf,Pomarol:1999ad,Gherghetta:2000qt,Davoudiasl:2000wi,Csaki:2002gy,Hewett:2002fe,Agashe:2003zs,Agashe:2007zd}
to allow all the SM fields (except the Higgs) to propagate in the
extra dimension. The SM particles are then the zero-modes of the 5D
fields and the profile of a SM fermion in the extra dimension can be
adjusted using a mass parameter. If  1st and 2nd generation fermion
profiles peak near the Planck brane then FCNC operators and PEW
corrections will be suppressed by scales $\gg \tev$.
Even with this arrangement it is estimated that the $g^1$, $W^1$ and $Z^1$ masses
must be larger than about $3\tev$ (see the summary in
\cite{Agashe:2007zd}). 

If the gauge bosons and fermions do not propagate in the bulk, then
the strongest limits on $\lphi$ come, via \Eq{monekk},  from the lower
bound placed by the LHC  on the first graviton KK excitation (see, for
example, \cite{ATLAS:2011ab}  and \cite{Chatrchyan:2011fq} for the
ATLAS and CMS limits). 
However, when the fermions propagate in the bulk, the couplings of
light fermion pairs to $G^1$ are greatly reduced and these limits do
not apply. 
When gauge bosons propagate in the bulk, a potentially important
experimental limit on the model comes from lower bounds on the 1st
excitation of the gluon, $g^1$.   In the model of
~\cite{Agashe:2006hk}, in which light fermion profiles peak near the
Planck brane, there is a universal component to the light quark
coupling $q\anti q g^1$ that is roughly equal to the SM $SU(2)$ gauge coupling $g$
times a factor of $\zeta^{-1}$, where $\zeta\sim \sqrt{\mbt} \sim
5-6$. The suppression is due to the fact that the light quarks are
localized near the Planck brane whereas the KK gluon is localized near
the TeV brane. Even with such suppression, the LHC $g^1$ production
rate due to $u\bar u$ and $d\bar d$ collisions is large. Further,
whatever the model, the $t_R\bar t_R g^1$ coupling is large since the
$t_R$ profile peaks near the TeV brane -- the prediction of
\cite{Agashe:2006hk} is $g_{t_R\bar t_R g^1}\sim \zeta g$.  As a
result, the dominant decay of the $g^1$ is to $t\bar t$. ATLAS and CMS
search for $t\bar t$ resonances at high mass. Using $g_{ q\bar q
  g^1}\sim g\zeta^{-1}$, $q=u,d$,  one finds a lower bound of  $m_1^g\gsim
1.5\tev$ \cite{Rappoccio:2011nj} using an update of the analysis of
\cite{Agashe:2006hk}.  (\cite{atlasgkk}  gives a weaker bound of
$m_1^g>0.84\tev$.) .  

In terms of $\lphi$, we have the following relations:
\beq
\frac{m_0}{\mpl}= \frac{\sqrt{6}}{x_1^g} \frac{m_1^g}{\lphi} \simeq \frac{m_1^g}{\lphi}\,,\quad \mbox{and}\quad{\mbt}=-\log \left( {\lphi\over \sqrt 6\mpl}\right)
\label{tie}
\eeq
where $x_1^g=2.45$ is the 1st zero of an appropriate Bessel function. 
If the model really solves the hierarchy problem then $\lphi \lsim 10\tev$ is required.
If we adopt the CMS limit of $m_1^g > 1.5\tev$ then \Eq{tie} implies a lower  limit on 
the 5-dimensional curvature of $\mompl \gsim 0.15$. Thus, a
significant lower bound on $m_1^g$ implies that only relatively  
large  values for $m_0/\mpl$  are allowed. As discussed above,
$\mompl$ values up to $\sim 2-3$ are probably consistent with
curvature corrections to the RS scenario being small. Still, tension
between the lower bound on $m_1^g$ and keeping acceptably small
$\mompl$ could increase to an unacceptable point  as the  LHC data set
increases.  We will discuss the phenomenology that applies if the
value of $\lphi$ for any given $(\mompl)$ is tied to the lower bound
of $m_1^g=1.5\tev$ using \Eq{tie}. Alterations to the phenomenology
using $m_1^g=3\tev$, as perhaps preferred by PEW constraints, will
also be illustrated. 

%
%
%

However, there are alternative approaches in which a lower bound on
$m_1^g$ from the LHC implies a less tight bound on $\lphi$. 
For example, including brane kinetic terms localized on the visible brane
for gauge fields and gravity will modify the Kaluza-Klein spectrum and the couplings of the fields
\cite{Davoudiasl:2002ua,Carena:2002dz,Davoudiasl:2003zt}. In
particular, the relation between  $\mompl$, ${m_1^g}$ and ${\lphi}$ will be 
modified in such a way that a large lower bound on ${m_1^g}$ can
still allow ${\lphi}$ sufficiently
low that the radion will have phenomenological impact. In this paper,
we thus also examine a non-minimal model in which no $\mompl$-dependent tie between
$m_1^g$ and $\lphi$ is assumed, implying that direct and indirect bounds on
${m_1^g}$ do not exclude the relatively low  values of $\lphi=1.5\tev$ and $1\tev$ for even relatively low values of $\mompl$.

Since the radion and Higgs fields have the same quantum numbers, it is
generically possible to introduce some amount of mixing between them.
When the Higgs is localized on the TeV brane, this mixing can be
introduced through an action operator that can be written in the
form~\cite{Giudice:2000av}:  
\beq
S_\xi=\xi \int d^4 x \sqrt{g_{\rm vis}}R(g_{\rm vis})\Hhat^\dagger \Hhat\,,
\label{grw}
\eeq
where $R(g_{\rm vis})$ is the Ricci scalar for the metric induced 
on the visible brane,
and 
$\Hhat$ is the Higgs field in the 5-D context before rescaling
to canonical normalization\footnote{Note however that in the case of a
Higgs leaking into the bulk, the 5D Higgs potential itself will induce
some mixing with the radion, which should be added to that coming
from Eq.~(\ref{grw}). For simplicity we will restrict ourselves to the case of a 
brane localized Higgs.}
.
The physical mass eigenstates, $h$ and $\phi$, are obtained by
diagonalizing and canonically normalizing the  kinetic (and mass) terms in the Higgs-radion Lagrangian.
The diagonalization procedures and results for  the $h$ and $\phi$  using our notation can be found
in~\cite{Dominici:2002jv} (see also
\cite{Giudice:2000av,Hewett:2002nk}).\footnote{In the current paper we
  change the sign of our convention for $\phi_0$.  We also note that
  in  \cite{Giudice:2000av,Hewett:2002nk} the coefficients in the
  $h_0$ decomposition are denoted by $a,b$ and those in the $\phi_0$
  decomposition are denoted $c,d$, \ie\ the reverse of our
  conventions.} 
One finds  
\beq
h_0=dh+c\phi\quad -\phi_0=a\phi +b h\,,\quad \mbox{where} \quad d=\cos\theta-t \sin\theta,~c=\sin\theta +t\cos\theta,~a=-{\cos\theta \over Z},~b={\sin\theta\over Z}\,,
\label{hphidef}
\eeq
with $t=6\xi \gam/ Z$, $Z^2=1+6\xi\gam^2(1-6\xi)$ and $\tan
2\theta=12\gam\xi Z
m_{h_0}^2/(m_{\phi_0}^2-m_{h_0}^2[Z^2-36\xi^2\gam^2])$. Here
$m_{h_0}^2$ and $m_{\phi_0}^2$ are the Higgs and radion masses before
mixing.  Consistency of the diagonalization imposes strong
restrictions on the possible $\xi$ values as a function of the final
eigenstate masses $\mh$ and $\mphi$, which restrictions depend
strongly on the ratio $\gam\equiv v_0/\lphi$ ($v_0=246\gev$).

The full Feynman rules after mixing for the $h$ and $\phi$
interactions with gauge bosons and fermions located in the bulk were
derived in \cite{Csaki:2007ns}. 
Of particular note are the anomaly terms associated with
the $\phi_0$ interactions before mixing. To be precise, we give a few
details of these important couplings and their implications. Let us
begin by defining 
\beq
\gh=(d+\gam b) \quad \gphi=(c+\gam a) \quad \ghr=\gam b\quad \gphir=\gam a\,.
\eeq
Relative to the Feynman rules of Fig.~29 of \cite{Dominici:2002jv},
the following modifications of the $gg$ and $\gam\gam$ couplings are
required when the gauge bosons propagate in the bulk: 
\bea
c_g^{h,\phi}&=&-{\alpha_s \over 4\pi v}\left[ g_{h,\phi}\sum_i F_{1/2}(\tau_i) - 2(b_3+{2\pi\over \alpha_s\mbt}) g^r_{h,\phi}\right]\cr
c_\gam^{h,\phi}&=&-{\alpha \over 2\pi v}\left[ g_{h,\phi} \sum_i e_i^2 N_c^i F_i(\tau_i) -(b_2+b_Y+{2\pi\over\alpha \mbt}) g^r_{h,\phi}\right]
\eea
(In Fig.~29 of \cite{Dominici:2002jv} we used the notation $g_{fV}$
for what we here call $g_{h,\phi}$. Also $g_r$ from
\cite{Dominici:2002jv} is replaced here by $g_{h,\phi}^r$ which  
incorporates the bulk propagation effects by the virtue of the second
term in the parentheses above). Since $b_3=7$ and $b_2+b_Y=-11/3$, the
new $g^r_{h,\phi}$ corrections can be significant.

There are also modifications to the $WW$ and $ZZ$ couplings of the $h$
and $\phi$ relative to Fig.~29 of \cite{Dominici:2002jv}.  Without
bulk propagation, these couplings were simply given by SM couplings
(proportional to the metric tensor $\eta^{\mu\nu}$) times $\gh$ or
$\gphi$.  For the bulk propagation case, there are additional terms in
the interaction Lagrangian that lead to Feynman rules that have terms
not proportional to $\eta^{\mu\nu}$, see \cite{Csaki:2007ns}. For
example, for the $W$ we have (before mixing) 
\beq
\call \ni h_0 {2\mw^2\over v}W_\mu^\dagger W^\mu -\phi_0 {2\mw^2\over \lphi} \left[ W_\mu^\dagger W^\mu\left(1-\kap_W\right) +W_{\mu\nu}^\dagger W^{\mu\nu}{1\over 4\mw^2 (\mbt)}\right]
\eeq
where $\kap_V=\left({3m_V^2(\mbt)\over\lphi^2(\mompl)^2}\right)$ for $V=W,Z$.  After mixing, this becomes, for example for the $h$ interaction
\beq
\call\ni h {2\mw^2\over v}\left[g_h^W W^\dagger_\mu W^\mu+g_h^r {1\over 4\mw^2 (\mbt)}W_{\mu\nu}^\dagger W^{\mu\nu}\right]\equiv h{2\mw^2\over v} g_h^W\left[W^\dagger_\mu W^\mu+\eta_h^W 
W_{\mu\nu}^\dagger W^{\mu\nu}\right]\
\eeq
with a similar result for the $\phi$. Here we have defined
\beq
g_{h,\phi}^V\equiv g_{h,\phi}-g^r_{h,\phi}\kap_V\,, \quad
\eta_{h,\phi}^V\equiv {g_{h,\phi}^r\over g_{h,\phi}^V} {1\over 4 m_V^2(\mbt)}\,.
\eeq
Substituting one $\mw=\half gv$ this gives the Feynman rule for the $hWW$ coupling as
\beq
ig\mw g_h^W\left[\eta_{\mu\nu} (1-2 k^+\cdot k^-\eta_{h}^W) +2 \eta_{h}^W k_\mu^+k_\nu^- \right]
\eeq
where $k^+,k^-$ are the momenta of the $W^+,W^-$, respectively.  The notations and results
for the $\phi$ and for $V=Z$ are obtained by corresponding modifications. Now, defining $R_{h,\phi}^V= 2\eta_{h,\phi}^Vm_V^2/(1-2 k^+\cdot k^-\eta_{h,\phi}^V)$ and  $x_V^{h,\phi}\equiv 4m_V^2/m_{h,\phi}^2$,
one finds that the matrix-element-squared for $h,\phi \to VV$ is proportional to
\beq
(g_{h,\phi}^V)^2(1-2k^+\cdot k^- \eta_{h,\phi}^V)^2\left\{\left[1-x_V^{h,\phi}+{3\over 4} (x_V^{h,\phi})^2\right]+R_{h,\phi}^V\left[-6+{4\over x_V^{h,\phi}}+2x_V^{h,\phi}\right]+(R_{h,\phi}^V)^2\left[4+{4\over (x_V^{h,\phi})^2}-{8\over x_V^{h,\phi}}\right]\right\}\,,
\eeq
where 
$k^+\cdot k^-=(2m_V^2/x_V^{h,\phi})(1-\half x_V^{h,\phi})$. 
The SM result would be obtained by setting $g_{h,\phi}^V=1$ and $\eta_{h,\phi}^V=0$.

In the case of fermions propagating in the bulk, both the radion and
the Higgs couplings to SM fermions can be  slightly modified. The couplings of the radion to  
the TeV-brane-localized top quark will receive no corrections with respect
to the original setup.  However, for quarks that are localized near the UV brane 
 (including the right-handed bottom), the modifications to
the radion quark couplings can be of order  
$\sim10\%-20$\% \cite{Csaki:2007ns}. Moreover, these coupling modifications are
not universal and so will also produce some amount of flavor violation into the
couplings of the radion with fermions \cite{Azatov:2008vm}. Observing
any of these effects will be challenging at the LHC and so in general we will neglect them.

Even though we neglect bulk effects in Yukawa couplings, it is worth commenting further on
the possible consequences of fermions propagating in the bulk. As an illustration, 
we briefly discuss the case of the unmixed Higgs interacting with
fermions that are allowed to propagate in the bulk.
The interaction term between the brane Higgs and the  up-type fermions can be written as
\beq
{ S}_{Y}= \int d^4xdy \sqrt{g_{vis}}\ \delta(y-y_{vis}) \left(H \bar{Q}_LY_1 U_R + 
H\bar{Q}_RY_2U_L + {\rm h.c}\right), \label{ 5DY}
\eeq
where $Y_1$ and $Y_2$ are $3\times 3$ complex matrices in flavor
space. For simplicity, we consider a setup in which the electroweak gauge symmetry
imposed on the model is that of the SM, \ie\ $SU(2) \times
U(1)$.
\footnote{In order to reduce tensions from PEW constraints one could consider extending the
  gauge symmetry group in order to add some built-in custodial
  symmetry protection (see \eg\ \cite{Agashe:2003zs}).}
The term $\delta(y-y_{vis}) H$ represents an $SU(2)$ Higgs doublet field localized
  on the visible brane, whereas $Q=Q_L+Q_R$ and $U=U_L+U_R$ are 5D
  fermion fields, transforming as doublet and singlet under $SU(2)$
  respectively. Note that in general  5D fermions have vectorlike representations,
  and in order to obtain a chiral low energy theory, one must
  impose vanishing boundary conditions (Dirichlet boundary conditions)
  on the field components $Q_R$ and $U_L$. Doing so eliminates these
  components from the lowest Kaluza Klein level, ensuring a chiral
  theory for the zero-mode fermions (which are therefore understood
 to be the SM fermions). The Yukawa operators in Eq.~(\ref{ 5DY}) are
localized on the visible brane, and are therefore 
chiral, \ie\ the left and right handed components of the  5D fermions
can be treated differently. Thus, we should generally consider  $Y_1\neq
Y_2$. 
In \cite{Azatov:2009na} it was shown that 
the operator proportional to $Y_2$ leads to the
appearance of flavor-violating couplings as well as potentially large corrections to
the diagonal Higgs Yukawa couplings of the effective theory.
These $Y_2$-operators can also potentially modify the radiative coupling of the Higgs to photons and, especially, to
gluons \cite{Casagrande:2010si,Azatov:2010pf} . 
The parametric dependence of these two effects (the corrections to the
Higgs-fermion couplings and the correction to the Higgs-gluon coupling) is
the same and goes as 
$Y_1Y_2^{\dagger} \frac{v^2}{M_{KK}^2}$, where $M_{KK}$ is the mass
scale of the KK fermions and $v$ is the Higgs vev. Perturbativity requires $|Y_1|,|Y_2|\lsim {\cal O}(3)$ \cite{Csaki:2008zd,Agashe:2008uz}. 
As the size of these  5D Yukawa couplings is reduced, so are the
corrections induced.
In fact, to successfully generate the SM fermion
masses only the $Y_1$ operator is needed,  \ie\ $Y_2$ terms are not
necessary. Thus, if we take  $|Y_2|\ll |Y_1|$ so as to avoid $Y_2$-induced flavor violating couplings then
large $Y_1$ values are possible with no corrections to the Higgs couplings.  Indeed, if $|Y_1|$ is as large as possible and $|Y_2|$ is small, this will reduce FCNC effects coming from the KK gluon excitations \cite{Csaki:2008zd,Agashe:2008uz} as well as those from Yukawa Higgs couplings.
In what follows, we adopt this limit and neglect the $Y_2$-induced corrections to Higgs couplings. 
\footnote{The corrections (either enhancement or suppression) to Yukawa couplings and Higgs production cross sections arising if fully general situations are considered, \ie\ by employing moderate to
large entries in  {\it both} matrices $Y_1$ and $Y_2$, can be of order tens of percent
\cite{Azatov:2009na,Azatov:2010pf}.} Finally, we note that corrections to the $h_0 \gam\gam$ couplings from KK $W$ excitations were shown in~\cite{Azatov:2010pf} to be $\lsim 5\%$ and will be neglected in our study.


With all this in mind, our goal here in this paper is to illustrate
the complexity of the phenomenology of the Higgs-radion system in the
context of LHC data. We will show in particular that an approximate
fit to the most prominent ``excesses'' in the Higgs search data can be
explained in the context of the model. Earlier papers on this topic
include \cite{deSandes:2011zs}, \cite{Barger:2011qn} (see also
\cite{Barger:2011hu}) and \cite{Cheung:2011nv}.

\vspace*{-.2in}
\section{LHC Excesses}
\label{res}
\vspace*{-.1in}

The Large Hadron Collider (LHC) data from the ATLAS~\cite{atlashiggs}
and CMS~\cite{cmshiggs} collaborations suggests the possibility of a
fairly Standard Model (SM) like Higgs boson with mass of order
$123-128\gev$. In particular, promising hints appear of a narrow
excess over background in the $\gam\gam$ and $ZZ\to 4\ell$ final
states with strong supporting evidence from the $WW\to\ell\nu\ell\nu$
mode. The ATLAS results suggest that the $\gam\gam$ and $4\ell$ rates
may be significantly enhanced with respect to the SM expectation at a
mass near $125\gev$. The CMS $\gam\gam$ rate is maximal for
$M_{\gam\gam} \sim 124\gev$ and also appears to be somewhat enhanced
with respect to the SM expectation. At this mass the CMS  signals in
other channels, including $\ell\nu\ell\nu$ and $4\ell$ are roughly
consistent with the expectation for a SM Higgs.  In addition, CMS data
shows excesses in the $4\ell$ rate near $120\gev$ (at which mass they
do not see a $\gam\gam$ excess) and in the $\gam\gam$ rates near
$137\gev$ (at which mass there is no $4\ell$ excess), but neither is
confirmed in the ATLAS data.   

One important point regards the $\ww\to\ell\nu\ell\nu$ final state.  The
signal for a scalar state of any given mass will be spread out into
many bins of a variable such as the transverse mass, $m_T$, as a
result of the missing energy carried by the neutrinos. Thus, if there
are two scalar states that have equal production cross section times
$WW$ branching ratio both may contribute but their contribution will
depend upon the analysis cuts applied.  
This contrasts with the $ZZ\to 4\ell$ channel (the only $ZZ$ channel
analyzed for scalar masses below $200\gev$) and the $\gam\gam$ channel
both of which have excellent mass resolution so that excesses should appear
centered on the scalar state masses. For this reason, we focus on these latter channels.

In the context of the Higgs-radion model, positive signals can only
arise for two masses. If more than two excesses were to ultimately emerge,
then a more complicated Higgs sector will be required than the single
$h_0$ case we study here. Certainly, one can consider 
including extra Higgs singlets or doublets. For the moment, we presume
that there are at most two excesses.  In this case, it is sufficient
to pursue the single Higgs plus radion model. 

We will consider three cases, labelled as ATLAS, CMSA and CMSB. We
quantify the excesses in terms of the best fit value for $R(X)\equiv
\sigma(X)/\sigma_{\rm SM}(X)$ for a given final state $X$. Errors
quoted for the excesses are those for $\pm 1\sigma$.  The mass
locations and excesses in the $\gam\gam$ and $4\ell$ channels in these
three cases, tabulated in Table~\ref{exc}, are 
taken from Figs.~8a and 8b of \cite{atlashiggs} in the ATLAS case and
from the appropriate windows of Fig.~14 of \cite{cmshiggs} 
in the case of CMSA and CMSB:
\vspace*{-.1in}
\begin{table}[h!]
\caption{Three scenarios for LHC excesses in the $\gam\gam$ and $4\ell$ final states.}
\begin{tabular}{|l|c|c|c|}
\hline
 & $125\gev$ (ATLAS) or $124\gev$ (CMS) & $ 120\gev$ & $137\gev$ \cr
 \hline
ATLAS & $R(\gam\gam)\sim 2.0^{+0.8}_{-0.8}$,  $R(4\ell)\sim 1.5^{+1.5}_{-1.0}$ & no excesses & no excesses \cr
\hline
CMSA & $R(\gam\gam)\sim 1.7^{+0.8}_{-0.7}$,  $R(4\ell)\sim 0.5^{+1.1}_{-0.7}$ & $R(4l)=2.0^{+1.5}_{-1.0}$,  $R(\gam\gam)<0.5$& no excesses \cr
\hline
CMSB & $R(\gam\gam)\sim 1.7^{+0.8}_{-0.7}$, $R(4\ell)\sim 0.5^{+1.1}_{-0.7}$ & no excesses & $R(\gam\gam)=1.5^{+0.8}_{-0.8}$, 
$R(4\ell)<0.2$ \cr 
\hline
\end{tabular}
\label{exc}
\end{table}
To an excellent approximation, only the $gg$ initial state is relevant for inclusive $h$ and $\phi$ production followed by decay to $\gam\gam$ or $ZZ\to 4\ell$ and so we will be comparing the ratios
\beq
R_h(X)\equiv {\Gamma_h(gg)\br(h\to X)\over \Gamma_{\hsm}(gg)\br(\hsm\to X)}\,,\quad R_\phi(X)\equiv {\Gamma_\phi(gg)\br(\phi\to X)\over \Gamma_{\hsm}(gg)\br(\hsm\to X)}\,,
\eeq
where numerator and denominator are computed for the same mass, to the
ATLAS, CMSA and CMSB $R(X)$ values quoted above. We also note that CMS
gives results for $W,Z+b\anti b$  relative to $W,Z+\hsm$ with $\hsm\to
b\anti b$ in the  SM at $120\gev$ and $124\gev$ of $1^{+1.4}_{-1.4}$
and $0.5^{+1.3}_{-1.5}$, respectively. No measurement for the $b\anti
b$ final state is quoted for $137\gev$. Finally, CMS has recently given results at $125\gev$ for the $\gam\gam$ final state in which the $WW$ fusion induced rate is separated from the $gg$ fusion induced rate \cite{moriondpas}. They find a ratio relative to the SM prediction for $WW\to\hsm\to \gam\gam$ of $R_{WW}(\gam\gam)=3.7^{+2.1}_{-1.8}$ at $125\gev$. Removing this $WW$ fusion component from the inclusive $\gam\gam$ final state gives a $gg$ fusion ratio of $R_{gg}(\gam\gam)=1.62\pm 0.69$. Were the $R_{WW}(\gam\gam)$ and $R_{gg}(\gam\gam)$ enhancements to both persist with increased statistics, it will be a huge challenge to the Higgs-radion approach (as we shall discuss) as well as to other models.

We note that the error bars on the SM multipliers for the ATLAS, CMSA
and CMSB scenarios are large and we regard it as likely that the
central values will surely change with more integrated luminosity at
the LHC. Increased integrated  luminosity  will hopefully  increase
the agreement between the ATLAS and CMS excesses, but could also
worsen the consistency, or perhaps even lead to the disappearance of
the excesses.  Thus, the comparisons below should only be taken as
illustrative of the possibilities. (Note that our plots are always
done with either $\mh$ or $\mphi$ equal to $125\gev$ as appropriate
for the ATLAS excess.  However, there is no change in the plots if we
use $124\gev$, as more precisely appropriate to the central value of
the CMSA and CMSB excesses.) 

As discussed above, it is appropriate to consider two different kinds
of models: a basic model in which a strong lower bound on the mass of the
first excited gluon implies a significant lower bound on $\lphi$ as a function of $\mompl$  and a model with non-minimal extensions such that a fixed (low) value of $\lphi$ can be considered for the full range of $\mompl$ even if there is a significant lower bound on $m_1^g$.
We consider these two alternatives in
turn. 

\subsection{\bfbm Lower bound on $m_1^g$}

In this section, we consider a model along the lines of
\cite{Agashe:2006hk} in which FCNC and PEW constraints are satisfied
by virtue of the fermionic profiles being peaked fairly close to the
Planck brane leading to fairly definitive couplings of the fermions to
the excited gauge bosons.  As described earlier, a lower bound of
$m_1^g\sim 1.5\tev$ can be obtained from LHC data while FCNC and PEW
constraints suggest a still higher bound of $\sim 3\tev$.  We will
show some results for both choices as we step through various possible
mass locations for the Higgs and radion that are motivated by the LHC
excesses in the $\gam\gam$ and/or $4\ell$ channels. In what follows,
each plot will be labelled by the value of $\mompl$ chosen and the
corresponding $\mpl\Omega_0$ value as  calculated for the fixed
$m_1^g$ using \Eq{tie}.  

\begin{figure}[h!]
  \bce
  \includegraphics[width=12cm,angle=90]{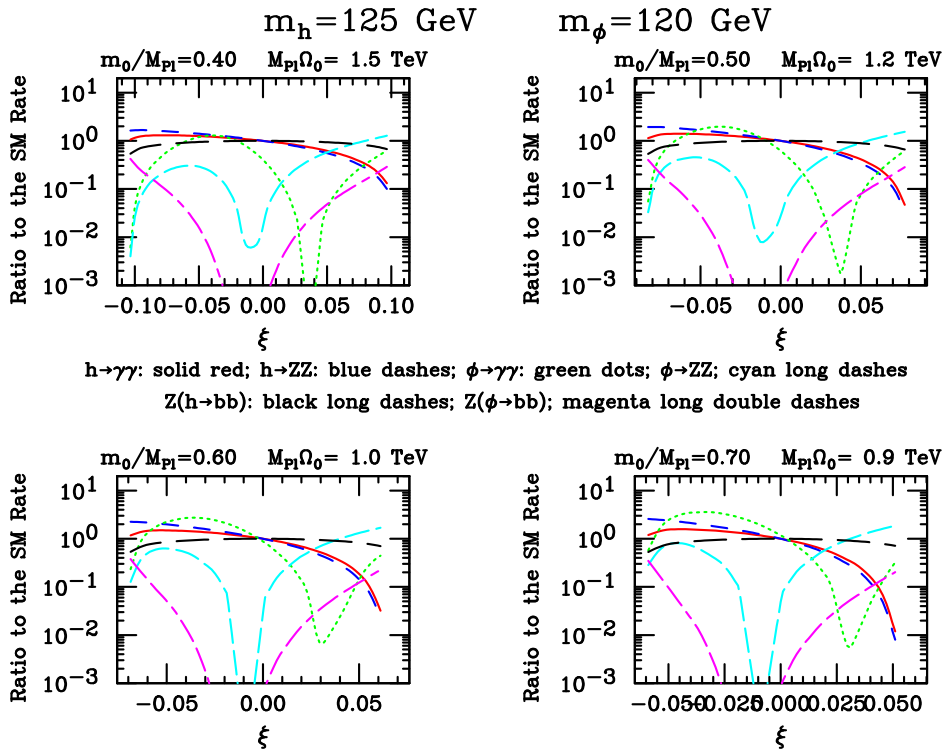}
\vspace*{-1.3cm}
\caption{For $\mh=125\gev$ and $\mphi=120\gev$, we plot $R_h(X)$ and $R_\phi(X)$  for $X=\gamma\gamma$ and $X=ZZ$ (equivalent to $X=4\ell$) as a function of $\xi$, assuming $m_1^g=1.5\tev$.  Also shown are the similarly defined ratios for $Z+h$ production with $h\to b\anti b$ and $Z+\phi$ production with $\phi\to b\anti b$.}
\label{mh125_mphi120_m1g1pt5tev_wbb_p4}
\ece
\vspace*{-.4in}
\end{figure}
\subsubsection{Signal at only {$125\gev$}}

In Fig.~\ref{mh125_mphi120_m1g1pt5tev_wbb_p4} we illustrate some possibilities for
$\mh=125\gev$ and $\mphi=120\gev$ taking $m_1^g=1.5\tev$.  First, we
note that to get an enhanced $\gam\gam$ rate at $125\gev$, it is
necessary to have $\mompl\gsim 0.4$ and $\xi<0$. In order to have
small $R_\phi(\gam\gam)$ and $R_\phi(4\ell)$ at $120\gev$ while at the
same time $R_h(\gam\gam)\gsim 1.5$ at $125\gev$, for consistency with
the ATLAS scenario, then $\mompl=0.4$ and $\xi \sim -0.09$ are good
choices.  The somewhat larger associated value of $R_h(4\ell)$ is
still consistent within errors with the ATLAS observation at
$125\gev$.  We note that for the reversed assignments of $\mh=120\gev$
and $\mphi=125\gev$, we cannot find parameter choices that yield a
decent description of the ATLAS $125\gev$ excesses with
$R_h(\gam\gam)$ and $R_h(4\ell)$  being sufficiently suppressed at
$120\gev$.  

\subsubsection {Signals at {  $125\gev$}  and {  $120\gev$}}

Fig.~\ref{mh125_mphi120_m1g1pt5tev_wbb_p4} also exemplifies the fact that with
$m_1^g=1.5\tev$ the Higgs-radion model is unable to describe the CMSA
scenario.  In the regions of $\xi$ for which appropriate signals are
present at $125\gev$ from the $h$, then at $120\gev$ the $4\ell$ and
$\gam\gam$ rates are either both suppressed or
$R_\phi(\gam\gam)>R_\phi(4\ell)$. This phenomenon persists at higher
$\mompl$ values as well as higher $m_1^g$.

\begin{figure}[p!]
  \bce
  \includegraphics[width=12cm,angle=90]{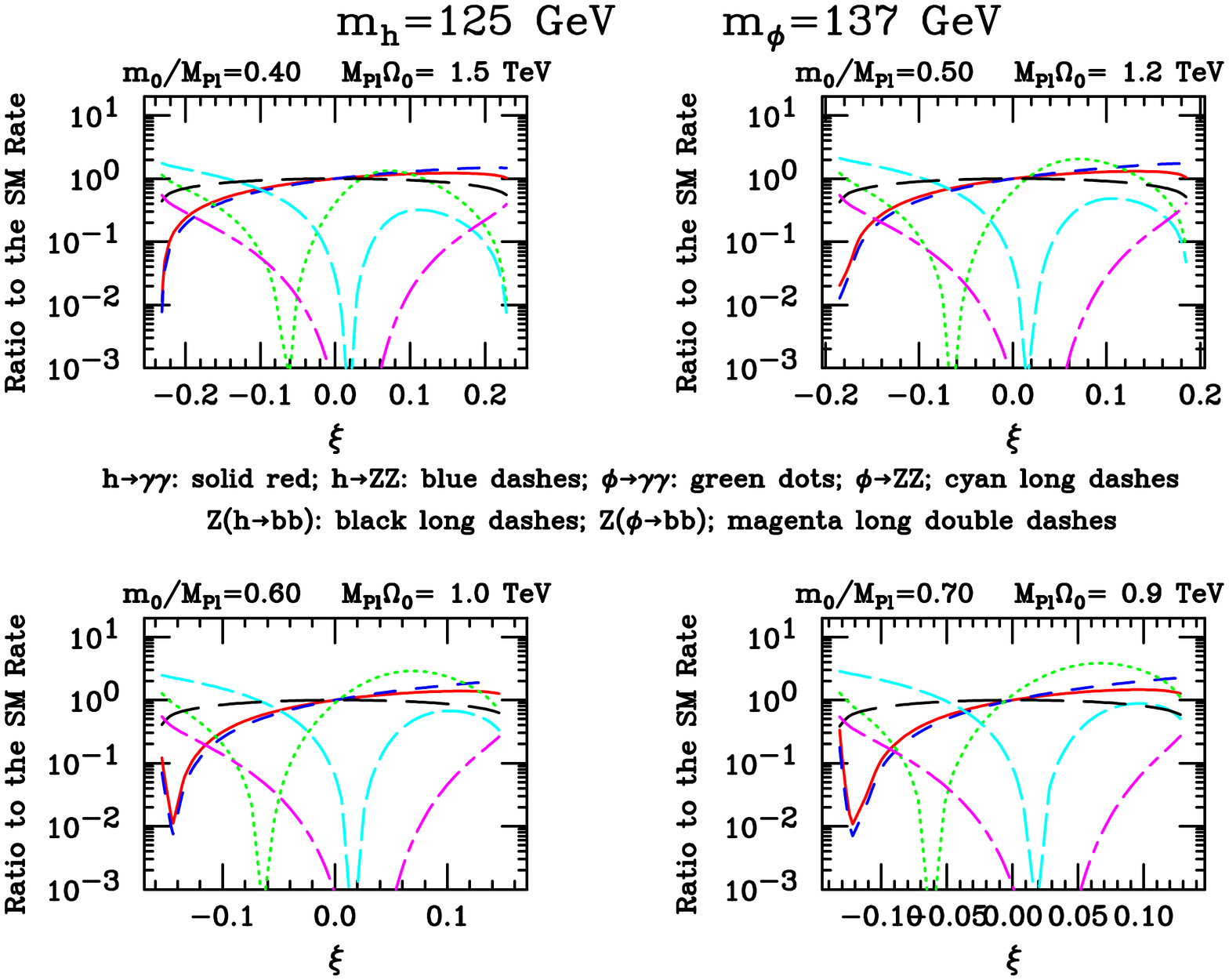}
\vspace*{-1.3cm}
\caption{For $\mh=125\gev$ and $\mphi=137\gev$, we plot $R_h(X)$ and $R_\phi(X)$  for $X=\gamma\gamma$ and $X=ZZ$ (equivalent to $X=4\ell$) as a function of $\xi$, assuming $m_1^g=1.5\tev$.   Also shown are the similarly defined ratios for $Z+h$ production with $h\to b\anti b$ and $Z+\phi$ production with $\phi\to b\anti b$.}
\label{mh125_mphi137_m1g1pt5tev_p4_withbb}
\ece
\vspace*{-.4in}
  \bce
  \includegraphics[width=12cm,angle=90]{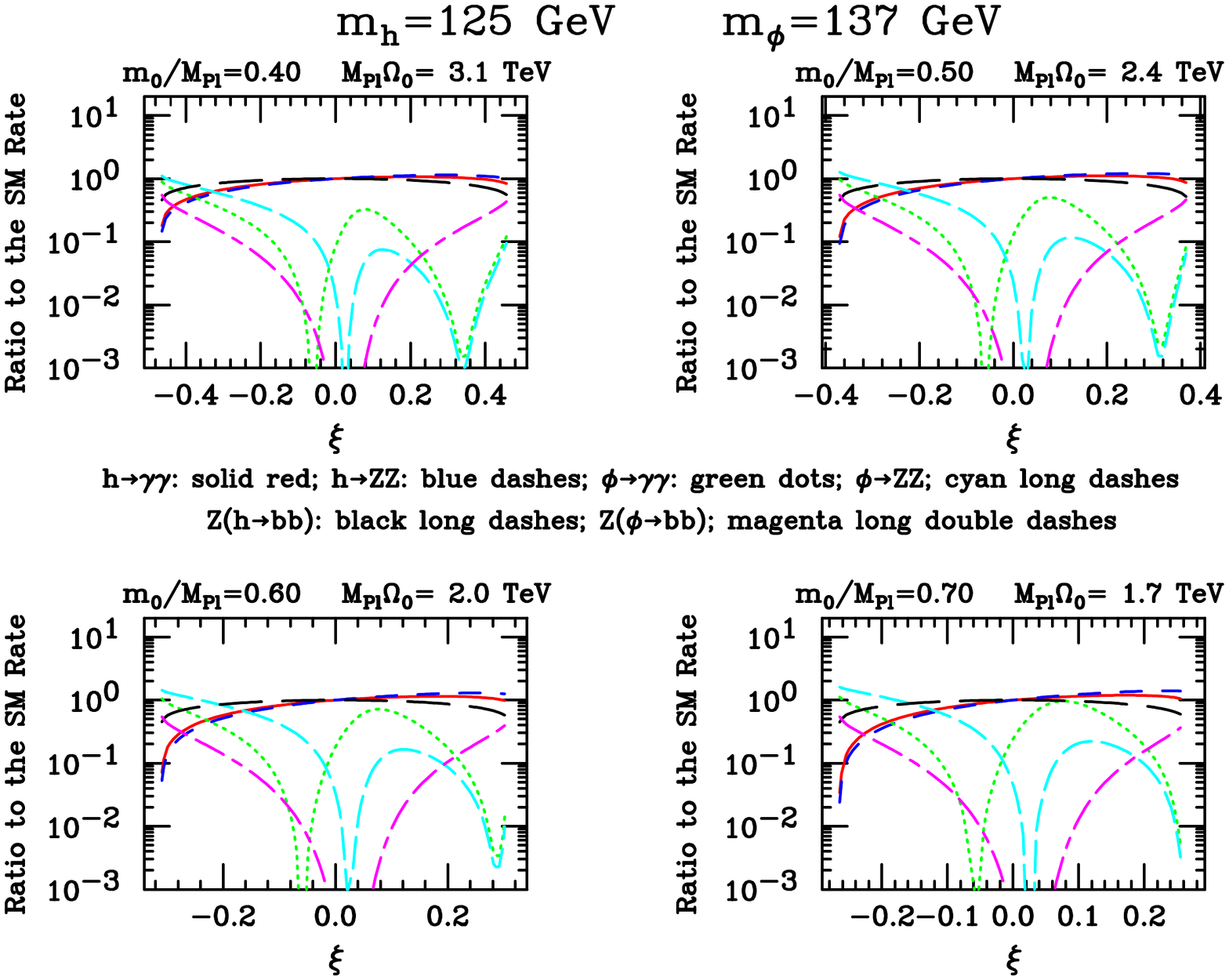}
\vspace*{-1.3cm}
\caption{For $\mh=125\gev$ and $\mphi=137\gev$, we plot $R_h(X)$ and $R_\phi(X)$  for $X=\gamma\gamma$ and $X=ZZ$ (equivalent to $X=4\ell$) as a function of $\xi$, assuming $m_1^g=3\tev$.  Also shown are the similarly defined ratios for $Z+h$ production with $h\to b\anti b$ and $Z+\phi$ production with $\phi\to b\anti b$.}
\label{mh125_mphi137_m1g3tev_wbb_p4}
\ece
\vspace*{-.2in}
\end{figure}

\subsubsection {Signals at {  $125\gev$}  and {  $137\gev$}}

Let us next consider the CMSB scenario, \ie\ neglecting the  $4\ell$
excess at $120\gev$ in the CMS data.  Taking $\mh=125\gev$ and $\mphi=137\gev$ with $m_1^g=1.5\tev$,
Fig.~\ref{mh125_mphi137_m1g1pt5tev_p4_withbb} shows that the choices
$\mompl=0.5$ and $\xi=0.12$ give $R_h(\gam\gam)\sim 1.3$ and
$R_h(4\ell)\sim 1.5$ at $125\gev$ and  $R_\phi(\gam\gam)\sim 1.3$ at
$137\gev$, fairly consistent with the CMSB observations. However,
$R_\phi(4\ell)\sim 0.5$ at $137\gev$ is a bit too large. Also shown in
the figure are the rates for $Z,W+h$ with $h\to b\anti b$ and
$Z,W+\phi$ with $\phi\to b\anti b$ relative to their SM
counterparts. For the above $\mompl=0.5$, $\xi=0.12$ choices, the
$Z,W+h(\to b\anti b)$ rate at $125\gev$ is only slightly below the SM
value, whereas the $Z,W+\phi(\to b\anti b)$ rate is about 10\% of the
SM level predicted at $137\gev$. The former is consistent with the
poorly measured $b\anti b$ rate at $124\gev$ while confirmation of the
latter would require much more integrated luminosity. 

We note that it is not possible to get enhanced $\gam\gam$ and $4\ell$ $h$ signals at
  $125\gev$ without having visible $137\gev$ $\phi$ signals, \ie\ the
ATLAS scenario of no observable excesses other than those at $125\gev$
cannot be realized for $\mphi=137\gev$. 
In addition, we note that for the $\mh=125\gev$ and $\mphi=137\gev$ mass assignment and $m_1^g=1.5\tev$, it is not possible to obtain  $R_{WW}(\gam\gam)$ significantly above 1.  More typically it is slightly below 1.
  
 For this case, it is also interesting to consider results for
 $\mh=125\gev$ and $\mphi=137\gev$ for the higher value of
 $m_1^g=3\tev$.  Results for this choice are plotted in
 Fig.~\ref{mh125_mphi137_m1g3tev_wbb_p4}.  We observe that $R_h(\gam\gam)$
 and $R_h(4\ell)$ are both $\lsim 1$ (or less) except for $\mompl=0.7$
 and large  $\xi$  
 for which  $R_\phi(\gam\gam)\ll 1$.  Thus, a reasonable description
 of the CMSB scenario requires relatively small $m_1^g$.  

Next, one can also consider the reversed mass assignments of $\mh=137\gev$ and $\mphi=125\gev$.
One finds that there is no choice of $\mompl$ at $m_1^g=1.5\tev$ for which the CMSB enhancements are approximately described. For $\xi$ choices for which  there is an enhanced $\gam\gam$ signal at $137\gev$, the $4\ell$ signal is even more enhanced.  One can find $\xi$ and $\mompl$ values such that the $\gam\gam$ and $4\ell$ signals are suppressed at $137\gev$ (\ie\ we seek a description of the ATLAS case) but for such choices there is no $\gam\gam$ enhancement at $125\gev$. As above, for $m_g^1=3\tev$ significant enhancements are not possible.

\subsubsection {Signals at   $125\gev$  and high mass }

 \begin{figure}[h!]
  \bce
  \includegraphics[width=12cm,angle=90]{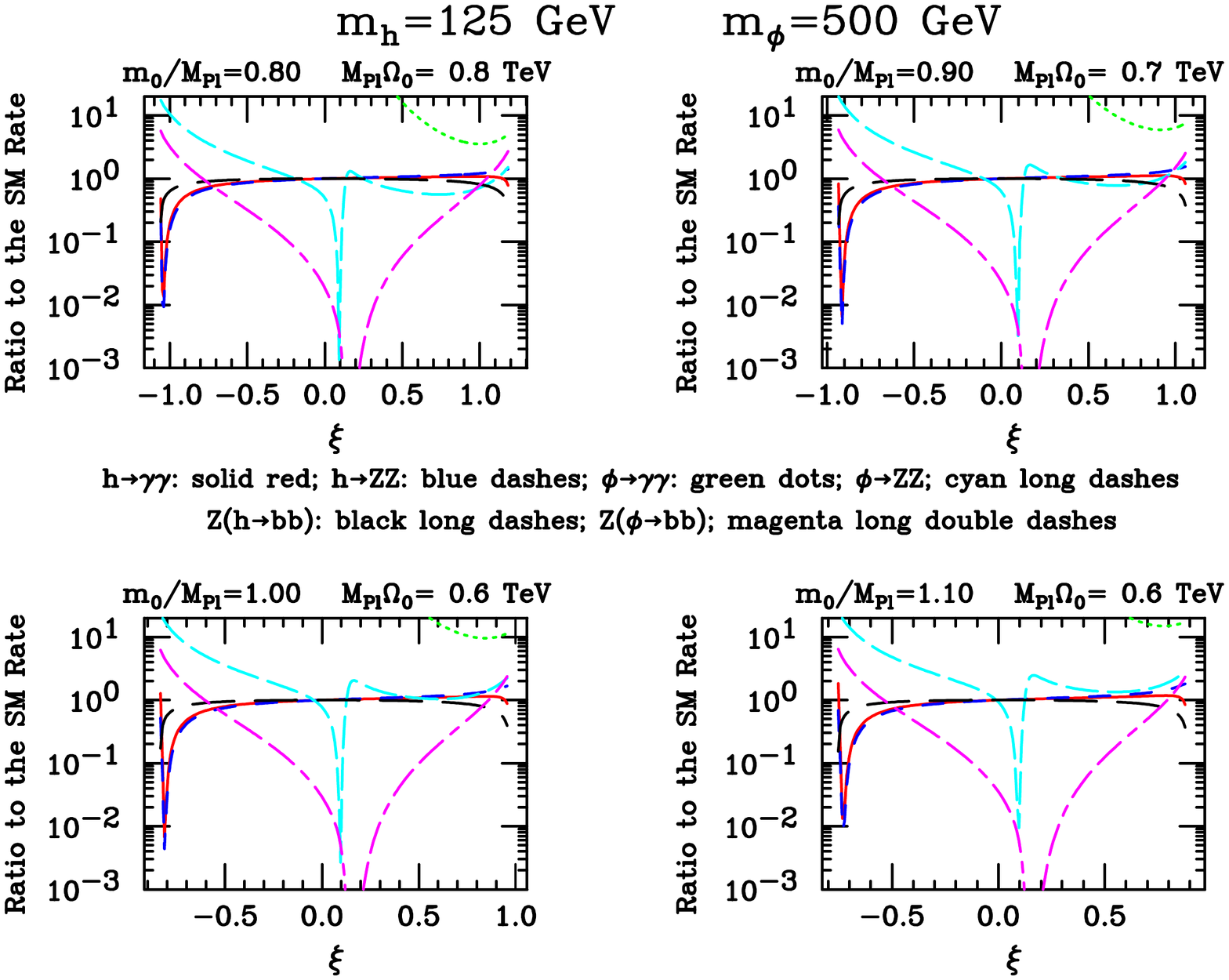}
\vspace*{-1.3cm}
\caption{For $\mh=125\gev$ and $\mphi=500\gev$, we plot $R_h(X)$ and
  $R_\phi(X)$  for $X=\gamma\gamma$ and $X=ZZ$ (equivalent to
  $X=4\ell$) as a function of $\xi$, assuming $m_1^g=1.5\tev$.  Also shown are the similarly defined ratios for $Z+h$ production with $h\to b\anti b$ and $Z+\phi$ production with $\phi\to b\anti b$.} 
\label{mh125_mphi500_m1g1pt5tev_wbb_p5}
\ece
\vspace*{-.2in}
\end{figure}

 A general question is whether one could explain the ATLAS $125\gev$
 excesses as being due to the $h$ or $\phi$ with the other being at
 high mass.  As shown in Fig.~\ref{mh125_mphi500_m1g1pt5tev_wbb_p5}, if $\mh=125\gev$ and $\mphi\sim 500\gev$, at $\mompl=1.1$ one finds 
 $R_h(\gam\gam)\sim 1.18$ and $R_h(4\ell)\sim 1.45$ for $\xi\sim 0.79$. As usual, the $4\ell$ signal is more enhanced (relative to the SM) than the $\gam\gam$ signal, but the above numbers are still consistent with the CMS $125\gev$ ratios within errors. For these same choices, the $\mphi=500\gev$ signal in the $4\ell$ final state would be of order 
 that expected for a SM Higgs at this same mass. CMS results in the $4\ell$ channel show a broad deficit in this same mass region that is inconsistent with the Higgs-radion prediction at the $2\sigma$ level.
 For the above parameter choices, the $\gam\gam$ signal at $\mphi=500\gev$ would be of order 8 times that for a SM Higgs at the same mass.

 Of course, it could happen that the CMS signals at $125\gev$ 
 drop to SM-level after more data is accumulated.  SM-like signals are obtained for $\mh=125\gev$ and $\mphi=500\gev$ at  moderate $\xi$ values. In this same parameter region, the 
heavy $\phi$ has a $4\ell$ rate that is suppressed relative to the SM, while the $\gam\gam$ rate is most typically highly enhanced, for example by a factor of $\sim 5000$ if $\xi\sim 0.1$ and $\mompl=1.1$.  If the $\gam\gam$ rate is  this large then the diphoton events 
 at  large invariant masses are likely to be observable~\cite{Toharia:2008tm}.  

Finally, we note that if $|\xi|$ is not modest in size when $\mphi$ is large, the $\phi VV$
($V=W,Z$) couplings become of SM strength or larger, thus adding
more pressure on the general setup coming from precision
electroweak constraints. For more discussion see \cite{Gunion:2003px}. 
 

 If the mass assignments are reversed, $\mh=500\gev$ and $\mphi=125\gev$, then 
 the $4\ell$ and/or $\gam\gam$ signals at $125\gev$ are suppressed relative to the SM. In addition,  this case is under tension from precision
 electroweak constraints since for all $\xi$ the $h$ alone 
 has $hVV$ couplings that are at least SM-like.  Much larger $\lphi$
 would be needed to have a hope of achieving PEW consistency from the
 Higgs-radion system \cite{Gunion:2003px}. In addition, the $h\to
 4\ell$ signal at high mass would be at least as large as predicted
 for a high-mass SM-like Higgs and therefore quite observable if
 $\mh\lsim 500\gev$, as seemingly inconsistent with ATLAS and CMS
 data.  If $\mh\sim 1\tev$, then the $4\ell$ signal would be beyond
 current LHC reach but PEW inconsistency would be much worse.

%
%

\subsection{Fixed $\lphi$}

In this section, we consider relaxing the tight relationship between $m_1^g$
and $\lphi$, which can occur in non-minimal scenarios as explained in
the introduction.   The relaxation of this relationship opens up additional phenomenological possibilities
as a result of the fact that one is then free to consider rather low values of $\lphi$ independent of $\mompl$ --- we will study $\lphi=1\tev$ and $\lphi=1.5\tev$, for which the Higgs-radion model
can yield LHC rates in the $\gam\gam$ and $4\ell$ channels that exceed
those that are predicted for a SM Higgs.  We note that when the gauge
bosons propagate in the bulk, the phenomenology does not depend on
$\lphi$ alone --- at fixed $\lphi$ explicit plots not given here show
that there is strong dependence on $\mompl$ when $\mompl$ is small.
However, for large $\mompl\gsim 0.5$ the phenomenology is determined
almost entirely by $\lphi$, but is still not the same as found in the
case where all fields are on the TeV brane. Once again, we step
through the various possible mass locations for the Higgs and radion
that are motivated by the LHC excesses in the $\gam\gam$ and/or
$4\ell$ channels. 

\vspace*{-.2in}
\subsubsection{Signal only at {$125\gev$}}

\begin{figure}[h!]
  \bce
  \includegraphics[width=12cm,angle=90]{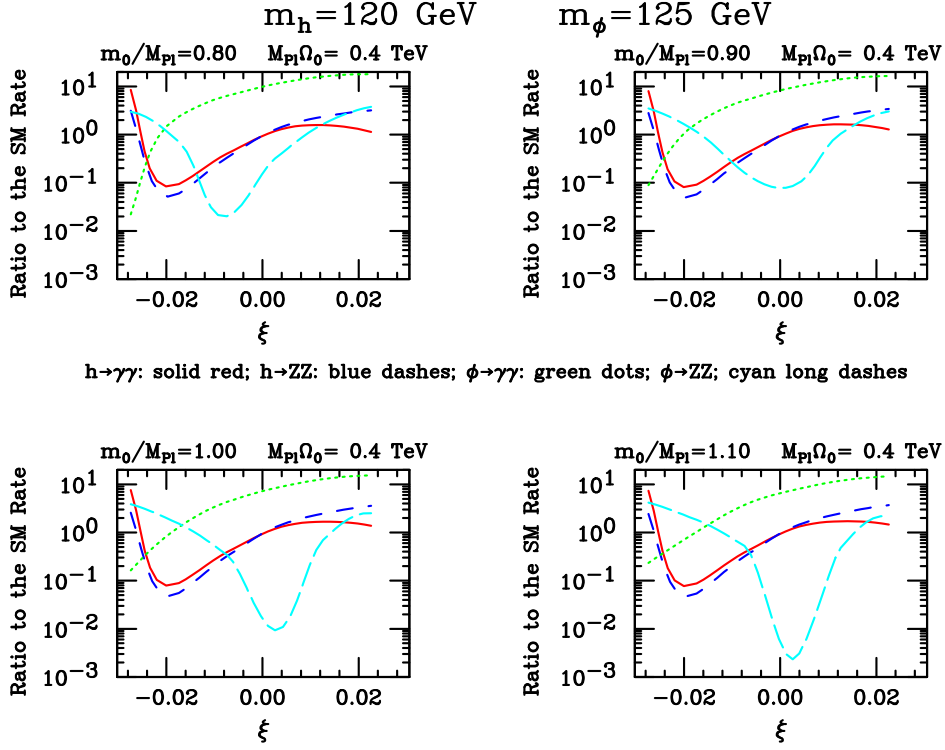}
\vspace*{-1.3cm}
\caption{For  $\mh=120\gev$ and $\mphi=125\gev$, we plot $R_h(X)$ and
  $R_\phi(X)$  for $X=\gamma\gamma$ and $X=ZZ$ (equivalent to
  $X=4\ell$) as a function of $\xi$ taking $\lphi$ fixed at $1\tev$.} 
\label{mh120_mphi125_lphi1tev_p4}
\ece
\vspace*{-.2in}
\end{figure}

As shown in Fig.~\ref{mh120_mphi125_lphi1tev_p4}, the choice of
$\lphi=1\tev$ with  $\mphi=125\gev$ and $\mh=120\gev$ gives a
reasonable description of the ATLAS excesses at $125\gev$ with no
visible signals at $120\gev$ in either the $\gam\gam$ or $4\ell$
channels when one chooses $\mompl=1$ and $\xi= -0.016$. In contrast,
for $\lphi=1.5\tev$ the $125\gev$ predicted excesses are below
$1\times$SM and thus would not provide a good description of the ATLAS
data. As exemplified in Fig.~\ref{mh125_mphi120_lphi1tev_p5},  
for the reversed assignments of $\mh=125\gev$ and $\mphi=120\gev$ any
choice of parameters that gives a good description of the $125\gev$
signals always yields a highly observable $120\gev$ signal. 

\subsubsection {Signals at {  $125\gev$}  and {  $120\gev$}}

\begin{figure}[h!]
  \bce
  \includegraphics[width=12cm,angle=90]{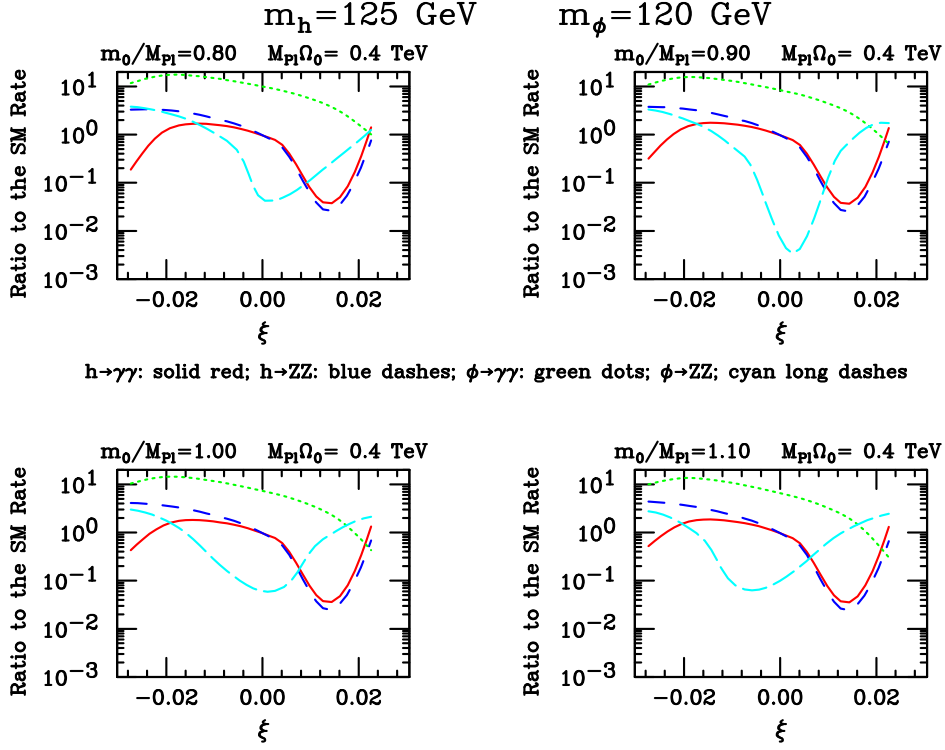}
\vspace*{-1.3cm}
\caption{For  $\mh=125\gev$ and $\mphi=120\gev$, we plot $R_h(X)$ and
  $R_\phi(X)$  for $X=\gamma\gamma$ and $X=ZZ$ (equivalent to
  $X=4\ell$) as a function of $\xi$ taking $\lphi$ fixed at $1\tev$.} 
\label{mh125_mphi120_lphi1tev_p5}
\ece
\vspace*{-.2in}
\end{figure}

We can also consider Fig.~\ref{mh125_mphi120_lphi1tev_p5} to see if
there is a choice of $\xi$ for which consistency with the CMSA
scenario is achieved. We observe that if $\xi$ is at its maximum value
and $\mompl=1.1$ then the $\gam\gam$ and $4\ell$ signals at
$\mh=125\gev$ are still within $-1\sigma$ of the CMS data while at
$\mphi=120\gev$ one finds  $R_\phi(4\ell)\sim 2.5$ while
$R_\phi(\gam\gam)\sim 0.3$, which values are roughly consistent with
the CMSA situation.  For the reversed assignments of $\mh=120\gev$ and
$\mphi=125\gev$, Fig~\ref{mh120_mphi125_lphi1tev_p4} illustrates the
fact that  a satisfactory description of the two CMSA excesses is not
possible --- for $\xi$ such that appropriate $125\gev$ excesses are
present, $R_h(\gam\gam)$ and $R_h(4\ell)$ at $120\gev$ are always
small so that the $4\ell$ excess at $120\gev$ is not explained.

\subsubsection {Signals at {  $125\gev$}  and {  $137\gev$}}


Let us now consider the CMSB scenario.  
%
For $\lphi=1\tev$, one finds $\mh=125\gev$ and $\mphi=137\gev$ with
the choices $\mompl=0.6$ and $\xi=-0.05$ give $R_h(\gam\gam)\sim 2$
and $R_h(4\ell)\sim 1$ at $125\gev$, while $R_\phi(\gam\gam)\sim 2$
and $R_\phi(4\ell)\sim 0.4$ at $137\gev$, an ok description of the
CMSB excesses.  An equally rough description of this same situation is
also possible for $\lphi=1\tev$ with $\mompl=0.8$ and $\xi=0.05$. 

\begin{figure}[h!]
  \bce
  \includegraphics[width=12cm,angle=90]{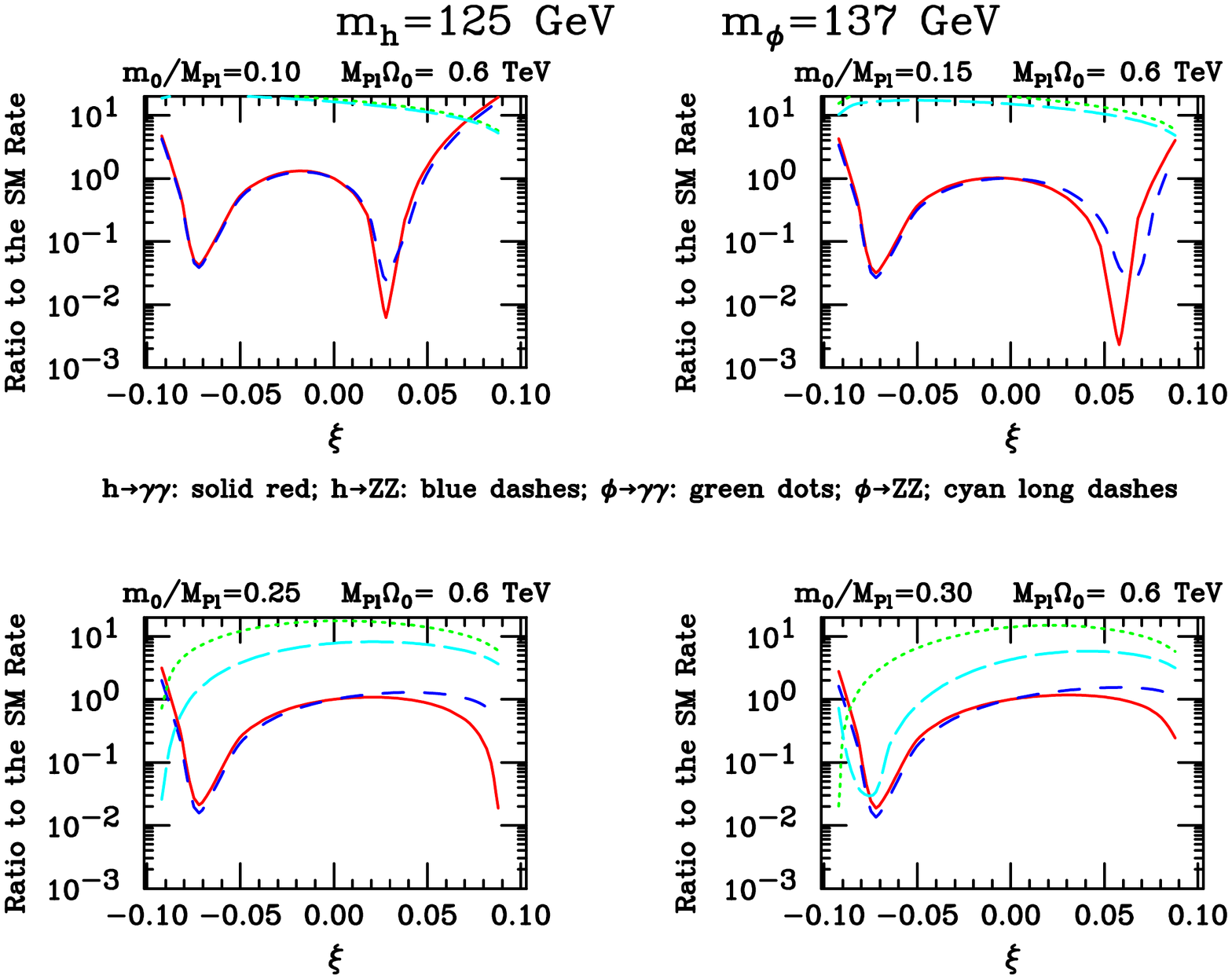}
\vspace*{-1.3cm}
\caption{For  $\mh=125\gev$ and $\mphi=137\gev$, we plot $R_h(X)$ and $R_\phi(X)$  for $X=\gamma\gamma$ and $X=ZZ$ (equivalent to $X=4\ell$) as a function of $\xi$ taking $\lphi$ fixed at $1.5\tev$.}
\label{mh125_mphi137_lphi1pt5tevf}
\ece
\vspace*{-.2in}
\end{figure}

For $\lphi=1.5\tev$ a somewhat better simultaneous description of
these excesses is possible. Fig.~\ref{mh125_mphi137_lphi1pt5tevf}
shows some results for  $\mh=125\gev$ and $\mphi=137\gev$. For
$\mompl=0.25$ and $\xi\sim -0.1$ one finds $R_h(\gam\gam)\sim 2$ and
$R_h(4\ell)\sim1.5$  at $\mh=125\gev$, while $R_\phi(\gam\gam)\sim 2$
and $R_\phi(4\ell)\ll 1$ at $\mphi=137\gev$, in pretty good agreement
with the CMSB scenario. 

If we reverse the configuration to $\mh=137\gev$ and $\mphi=125\gev$, only $\lphi=1\tev$ with $\mompl=0.8$ and $\xi\sim 0.05$ 
comes close to describing the two
excess; one finds that the $\mphi=125\gev$ $\gam\gam$ and $4\ell$
signals and the $\mh=137\gev$ $\gam\gam$ signal are all at the level
of $\sim 1.4\times$SM.  However, the $\mh=137\gev$ $4\ell$ signal is
at the level of $\sim 0.6\times$ SM which is $4\sigma$ away from the
CMS central value at this mass. For these mass assignments and the higher
$\lphi=1.5\tev$ value, $\mompl$ and $\xi$ choices that approximately describe the CMS excesses cannot be found--- the $\mphi=125\gev$ signals
are never simultaneously sufficiently large  to fit the
observed signals.

\subsubsection {Signals at {  $125\gev$}  and higher mass}

We choose not to show any specific plots for this situation.  For
$\lphi=1\tev$ or $1.5\tev$, it is possible to choose one of either the
$h$ or $\phi$ to have a mass of $125\gev$ and find $\mompl$ and $\xi$
values that result in a decent description of the $125\gev$ ATLAS
excesses.
When the $\phi$ is heavy, the scenario can be viable but the $\phi$
might be hard to dicover due to suppressed couplings to ZZ.
When the $h$ is heavy there would be tensions coming PEW constraints
and, if the higher mass is chosen below $500\gev$, a highly observable
$4\ell$ signal that would be inconsistent with ATLAS and CMS
observations in that region of mass.

\subsubsection{SM Higgs at $125\gev$ and Signal at $137\gev$}

\begin{figure}[h!]
  \bce
  \includegraphics[width=12cm,angle=90]{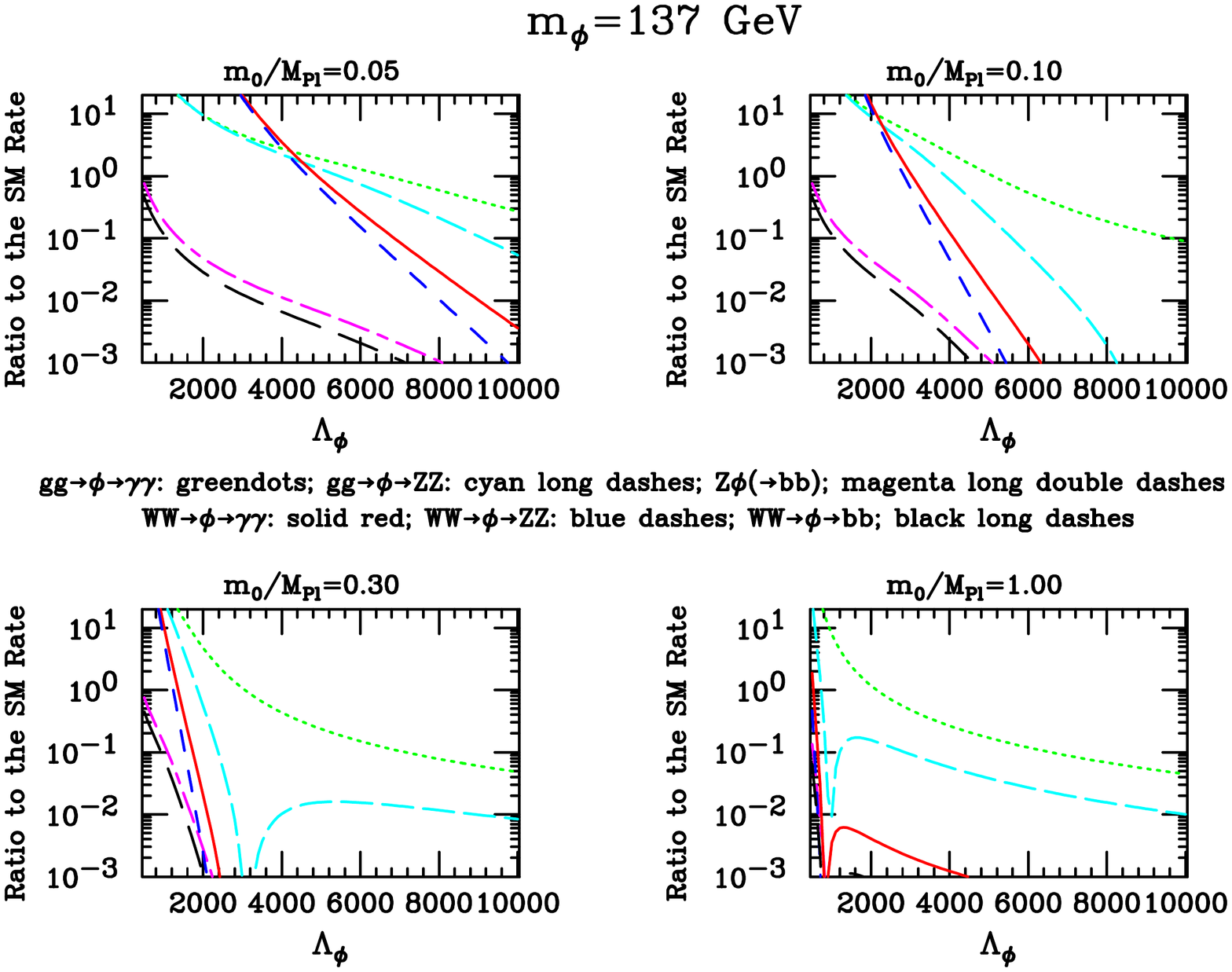}
\vspace*{-1.3cm}
\caption{For  $\mphi=137\gev$, we plot $R_\phi(X)$  for $X=\gamma\gamma$ and $X=ZZ$ (equivalent to $X=4\ell$) as  functions of $\lphi$ taking $\xi=0$. We also plot ratios to the SM for $Z\to Z\phi$ with $\phi\to b\anti b$ and for $WW\to \phi\to X$ for $X=\gam\gam$, $ZZ$ and $b\anti b$. }
\label{mh125_mphi137_lphiscanpaper}
\ece
\vspace*{-.2in}
\end{figure}

It is still quite conceivable that, after accumulating more data, the
excesses at $\sim 125\gev$ will converge to those appropriate for a SM
Higgs boson.  Such a situation would correspond to taking $\xi=0$ in
the Higgs-radion model.  In this case, one can ask whether or not
there could be a radion at some nearby mass and what its experimental
signature would be.  To exemplify, let us suppose that the signal at
$137\gev$ of the CMSB scenario survives.  In
Fig.~\ref{mh125_mphi137_lphiscanpaper} we plot 
$R_\phi(X)$ for $X=\gam\gam$ and $X=4\ell$ as a function of $\lphi$
for a selection of $\mompl$ values, taking $\xi=0$.   Also shown are ratios to the SM for $Z\to Z\phi$ with $\phi\to b\anti b$ and for $WW\to \phi\to X$ for $X=\gam\gam$, $ZZ$ and $b\anti b$. 
One observes that a nice description of the $R(\gam\gam)\sim 2$ excess
at $137\gev$ is possible, for example,  for $\mompl=0.3$ at $\lphi\sim
2.8\tev$ with the $4\ell$ signal (and all other signals)  being very suppressed.  As also
apparent, other choices of the $\mompl$ and $\lphi$ will also yield
$R_\phi(\gam\gam)\sim 2$ with varying levels of $4\ell$ and other
signals. (However, to suppress $R_\phi(4\ell)$ below $0.2$ while
achieving $R_\phi(\gam\gam)\sim 2$ requires $\mompl\geq 0.3$.)  
We also note that for $\xi=0$ the $Z,W+\phi(\to b\anti b)$ rates are greatly suppressed relative to their SM counterparts.

Plots for the case of a SM Higgs at $125\gev$ and $\mphi=120\gev$ look
very similar and, in particular, it is not possible to find parameters
for which the $4\ell$ signal substantially exceeds the $\gam\gam$
signal --- the reverse always applies, as one anticipates from the
enhanced anomalous $\gam\gam$ coupling of the (unmixed) $\phi$. 

%
\vspace*{-.1in}
\section{Summary and Conclusions}
\label{sum}
\vspace*{-.1in}

The Randall Sundrum model solution to the hierarchy problem yields
interesting phenomenology for the Higgs-radion sector, especially when
Higgs-radion mixing is allowed for, and can be made consistent with
FCNC and PEW constraints if fermions and gauge bosons propagate in the
5th dimension. At the moment, there are interesting hints at the LHC
of narrow excesses above SM backgrounds in the $\gam\gam$ and $ZZ\to
4\ell$ channels, as well as a broad excess in the $WW\to
\ell\nu\ell\nu$ channel.  ATLAS sees excesses in the $\gam\gam$ and
$4\ell$ channels at a mass of $\sim 125\gev$ of order $2\times$SM and
$1.5\times$SM respectively. CMS sees a $\gam\gam$ excess of order
$1.5\times$ SM at $\sim 124\gev$ and constrains the $4\ell$ channel at
this mass to be less than $\sim 1.5\times$SM.  Additional excesses at
$120\gev$ (in the $4\ell$ channel) and at $137\gev$ (in the $\gam\gam$
channel) are present in the CMS data.   

In this paper, we explored a wide range of possibilities within the
Randall Sundrum model context. In a first set of plots, we assumed the standard relation between 
$\lphi$ (the radion field vacuum expectation value), the curvature
ratio $\mompl$ and $m_1^g$ (the mass of
the 1st excited gluon state) that applies in the class of scenarios in which the 5th dimension profiles
for light fermions need to be peaked near the Planck brane in order to
avoid corrections to FCNC and PEW constraints that are too large.  We considered lower bounds on the latter of $m_1^g>1.5\tev$ or $3\tev$, as estimated using LHC
data.
Our second set of plots are done holding $\lphi$ fixed at either $1\tev$
or $1.5\tev$, using the fact that the lower bounds (as a function of $\mompl$) on $\lphi$ resulting from a lower bound on the mass of the $g^1$ can be
loosened in non-minimal extensions of the setup.   Our studies are done assuming that the Yukawa coupling of the brane Higgs to the  5D fermionic fields  proportional to $
H\bar{Q}_RY_2U_L + {\rm h.c.}$ is small.  Such a choice is consistent with FCNC and PEW constraints. In this case, the unmixed $h_0$ couplings, in particular to $gg$ and $\gam\gam$, are not modified with respect to those of a SM Higgs boson.  In this way, we sample an interesting range of phenomenological possibilities. For fully general $Y_2$, corrections to the $h_0$ couplings due to  5D effects can be large and can either suppress or enhance the $gg$ and $\gam\gam$ couplings by tens of percent. Even without this freedom, the mixed Higgs-radion phenomenology is quite diverse as we have shown. 

Since the single Higgs plus radion model can describe at most two
Higgs-like excesses, we considered three scenarios labelled as: ATLAS,
with $\gam\gam$ and $4\ell$ excesses at $125\gev$ larger than SM and
no other significant excesses; CMSA, with $\gam\gam$ and $4\ell$
excesses at $124\gev$ above SM level and a $4\ell$ excess at
$120\gev$; and, CMSB, with  $\gam\gam$ and $4\ell$ excesses at
$124\gev$ above those predicted for a SM Higgs boson of this same mass
along with a $\gam\gam$ excess at $137\gev$ that is also larger than
would have been predicted for $\mhsm=137\gev$.  In both the fixed
$m_1^g=1.5\tev$ and the fixed $\lphi=1\tev$ model possibilities, the signal levels
of the ATLAS and CMSB scenarios could be nicely described, whereas the enhancements relative to the SM were too small for $m_1^g=3\tev$ and $\lphi=1.5\tev$, respectively. A satisfactory description of the 
CMSA scenario was also found  in the case of fixed $\lphi= 1\tev$, but not in the case where fixed $m_1^g=1.5\tev$ was used to determine $\lphi$ as a function of $\mompl$. In general,
successful fitting of the ATLAS and CMSB excesses required a modest
value for the radion vacuum expectation value, mostly $\Lambda_\phi\lsim 1$ TeV, and typically $m_0/m_{Pl}\gsim 0.5$,
 a range that the most recent discussion suggests is
still consistent with higher curvature corrections to the RS scenario
being small.  

We also considered expectations for the radion signal in the case
where the Higgs signal was assumed to ultimately converge to precisely
that for a SM Higgs of mass $125\gev$.  This situation would arise if
there is no Higgs-radion mixing. We found that interesting
excesses at the radion mass would be present for low enough $\lphi$,
namely $\lphi\lsim 3\tev$, but would always be characterized by a
$\gam\gam$ signal that substantially exceeds the $4\ell$ signal (as
appropriate for the CMS $\gam\gam$ excess at $137\gev$ but in definite
contradiction with the CMS $4\ell$ excess at $120\gev$). Finally, we noted that a larger-than-SM signal in $WW\to h~{\rm or}~\phi\to \gam\gam$, as possibly seen by CMS at $125\gev$, cannot be achieved (at least in the model employed here where the unmixed $h_0$ couplings are SM-like).

%
\vskip .05in
\centerline{\bf Acknowledgments}

We thank Kaustubh Agashe, Felix Brummer, Csaba Csaki, Gero
Von Gersdorff, Mariana Frank, Tom Rizzo and John Terning for illuminating discussions.
BG is partially supported by the National Science Centre (Poland)
as a research project, decision no DEC-2011/01/B/ST2/00438. JFG is supported by US DOE grant DE-FG03-91ER40674.

\end{document}